\documentclass[aps,prb,reprint,amsmath,amssymb,superscriptaddress,floatfix,preprintnumbers]{revtex4-1}
\usepackage[english]{babel}
\usepackage{hyperref}
\usepackage{comment}
\usepackage{graphicx}
\usepackage{dcolumn}
\usepackage{bm}
\usepackage{color}
\newcommand{\eg}{{\em e.g.}\ }
\newcommand{\ie}{{\em i.e.}\ }
\newcommand{\etal}{{\em et al.}\ }
\newcommand{\figref}[1]{Fig.~\ref{#1}}
\newcommand{\secref}[1]{Section~\ref{#1}}
\newcommand{\buck}{$\text{C}_\text{60}$}
\begin{document}
\title{Electronic and transport properties of kinked graphene}
\author{Jesper Toft Rasmussen}
\affiliation{Center for Nanostructured Graphene (CNG), \\Department of Micro- and Nanotechnology (DTU Nanotech), \\Technical University of Denmark, DK-2800 Kongens Lyngby, Denmark}
\author{Tue Gunst}
\affiliation{Center for Nanostructured Graphene (CNG), \\Department of Micro- and Nanotechnology (DTU Nanotech), \\Technical University of Denmark, DK-2800 Kongens Lyngby, Denmark}
\author{Peter B{\o}ggild}
\affiliation{Center for Nanostructured Graphene (CNG), \\Department of Micro- and Nanotechnology (DTU Nanotech), \\Technical University of Denmark, DK-2800 Kongens Lyngby, Denmark}
\author{Antti-Pekka Jauho}
\affiliation{Center for Nanostructured Graphene (CNG), \\Department of Micro- and Nanotechnology (DTU Nanotech), \\Technical University of Denmark, DK-2800 Kongens Lyngby, Denmark}
\author{Mads Brandbyge}
\email{Mads.Brandbyge@nanotech.dtu.dk}
\affiliation{Center for Nanostructured Graphene (CNG), \\Department of Micro- and Nanotechnology (DTU Nanotech), \\Technical University of Denmark, DK-2800 Kongens Lyngby, Denmark}

\begin{abstract}
Local curvature, or bending, of a graphene sheet is known to increase the chemical reactivity presenting an opportunity for templated chemical functionalization. Using first principles calculations based on density functional theory (DFT) we investigate the reduction of the reaction barrier for adsorption of atomic hydrogen at linear bends in graphene. We find a significant lowering ($\sim\!15$\%) for realistic radii of curvature ($\sim\!20$~\AA), and that adsorption along the linear bend leads to a stable linear kink. We compute the electronic transport properties of individual and multiple kink-lines, and demonstrate how these act as efficient barriers for electron transport. In particular, two parallel kink-lines form a graphene pseudo-nanoribbon structure with a semi-metallic/semi-conducting electronic structure closely related to the corresponding isolated ribbons; the ribbon band gap translates into a transport gap for transport across the kink lines. We finally consider pseudo-ribbon based heterostructures, and propose that such structures present a novel approach for band gap engineering in nanostructured graphene.
\end{abstract}

\date{\today}
\maketitle

\section{\label{sec:introduction}Introduction}

Nanostructures based on graphene have an enormous potential for applications. Especially in future electronic devices compatible with and extending silicon-technology, due to the outstanding electronic transport properties of graphene\cite{NoFaCo.12}. However, it is crucial to modify the semi-metallic electronic structure of graphene to exploit its full potential for many electronic applications: a band gap can be introduced by nanostructuring graphene.

A common approach towards engineering the electronic structure is to form quasi-1D graphene in the form of nanoribbons (GNR)\cite{HaAZh.07}. The electronic structure of GNRs depends on width, direction and edge structure -- all parameters which to some degree can be controlled. GNRs can be formed by etching\cite{HaAZh.07}, by unzipping carbon nano-tubes (CNTs)\cite{xie_graphene_2012}, or, ultimately be grown with atomic-scale precision using self-assembly of precursor molecules on metal substrates\cite{CaRuJa.2010}. However, for electronic applications this approach requires a structure-preserving means of releasing and transferring the structures to an insulating substrate. Bonding of H or other species to graphene with large coverage opens an insolating band gap at the adsorption sites due to $sp^3$ hybridization\cite{elias_control_2009}. Periodically ordered clusters of adsorbed hydrogen can be formed on graphene in patterns dictated by the Moir\'e lattice mismatch between graphene and the metal substrate, which opens a semi-conducting band gap\cite{BaJoNi.2010}. Finally, regular perforations, known as a graphene anti-dot lattice (GAL)\cite{Pedersen2008} or a nanoscale mesh of holes\cite{kisaha.2010,lijuwu.2010,bazhji.2010} can have neck widths\cite{bazhji.2010,shimizu_023104} down to 5~nm corresponding to band-gaps of the order of 1~eV\cite{HaAZh.07}.

\begin{figure}[tbp]
	\centering
	\includegraphics[width=.47\textwidth]{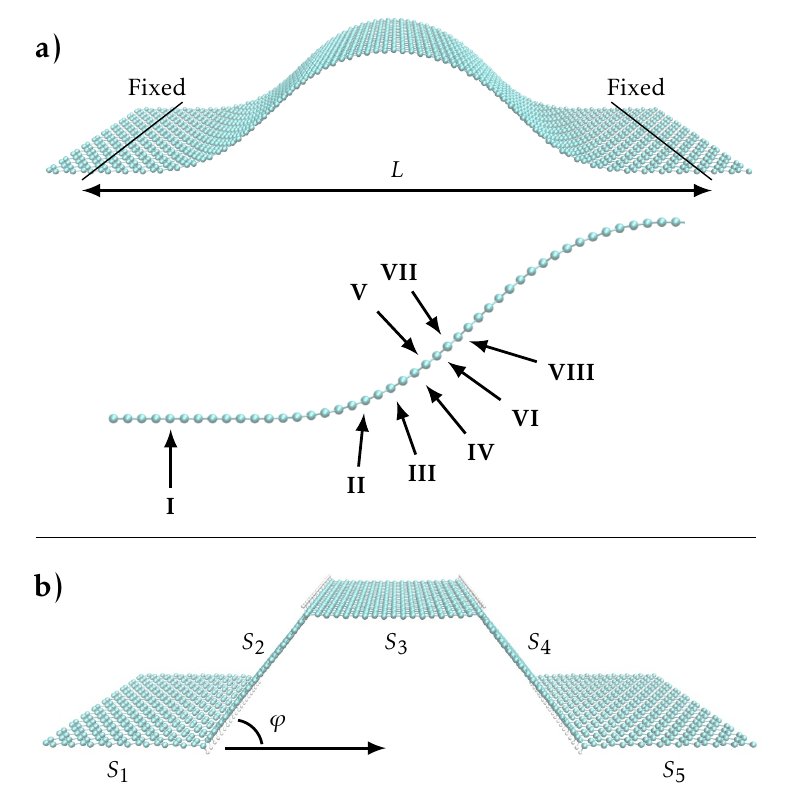}
	\caption{(Color online) (a) Smooth ripple-like structure where first and last six rows of carbon-dimers are surface-clamped regions with a separation of $L\!=90$~\AA. Atomic hydrogen is adsorbed at positions I--VIII. (b) The resulting kinked graphene structure after hydrogen is adsorbed in lines at the most reactive position (II) corresponding to the smallest local radius of curvature. The four kinks divide the structure into five sections, $S_1$--$S_5$.}
	\label{fig:figure1}
\end{figure}

Graphene consists entirely of surface atoms and is thus exceedingly sensitive to the surroundings. In particular, the van der Waals (vdW) interaction with the substrate is of importance. The substrate interactions, which make graphene cling to small features, may be exploited by manufacturing nanostructures in the substrate. Periodic steps in a Cu substrate has been used to induce "wrinkles" or ripples in graphene with period and height on the order of 10~nm\cite{pan_wrinkle_2011}.  Recently, Hicks \etal\cite{hicks_wide_2012} demonstrated how arrays of 1D large-bandgap, semiconducting graphene nanoribbons corresponding to a width of $\sim\!\!1.4$~nm can be formed in graphene on a step-patterned SiC substrate. The substrate interactions can clamp a graphene sheet while partly suspended across small holes, so that a pressure difference on the in- and outside lead to the formation of bubbles or "blisters" in the sheet\cite{Bunch2012}. Also, linear folds, where the graphene sheet is bulging up from the substrate, have been induced for graphene suspended over trenches using heat treatment\cite{bao_controlled_2009}. Thus, the sheet can obtain significant bends at certain places induced by the substrate interaction, substrate nanostructuring, and subsequent treatments\cite{PhysRevB.85.195445}. Calculations by Low \etal\cite{LoPeTe.2012} showed how a sharp step of height 1~nm in a SiC substrate, comparable to experimental values\cite{hicks_wide_2012}, can induce a linear bend in the graphene sheet with a radius of curvature down to around 1~nm.

The ability to accurately control sharp local curvatures of graphene presents opportunities for strain-assisted modification of the local electronic structure and chemical reactivity in the graphene sheet, and may open a route to band gap engineered devices\cite{lebume.2010,gukage.2010,hicks_wide_2012,ortolani_folded_2012,wu_selective_2012}. Very recently Wu \etal\cite{wu_selective_2012} showed how graphene on a Si substrate decorated with SiO$_2$ nanoparticles induced local regions of strain and increased reactivity in a selective manner. Atomic hydrogen or other chemical species do not easily react with flat graphene when dosed from a single side\cite{hornekaer_clustering_2006}. However, at positions where there is a substantial local bending, rippling or strain of the graphene sheet the reactivity changes significantly\cite{srivastava_predictions_1999,elias_control_2009}. So far there has been only a few theoretical studies of the atomic geometry of hydrogenated ripple structures in unsupported, strain-induced, graphene ripples\cite{WaZhLi.2011,chernozatonskii_nanoengineering_2010,chernozatonskii_two-dimensional_2007}. However, to the best of our knowledge no studies have addressed the reactivity of bends or the transport through hydrogenated ripples, or discussed the possibility of stabilizing non-planar structures by hydrogenation.

In this paper we consider the reactivity of linear bends in a graphene sheet, and the electronic transport properties of kinks resulting from hydrogenation of bends. Our starting point is the generic graphene structure shown in \figref{fig:figure1}a which is inspired by the experimental observation of trench formation\cite{bao_controlled_2009}. The bulging of this structure results from shortening the distance between two separated, clamped regions in the sheet. The remaining sections of the paper are organized as follows: In \secref{sec:setup} we describe our computational method and setup. In \secref{sec:results} we present our results: First we describe the adsorption barriers for reacting with single atomic H on the graphene bend at positions with different local curvature (positions I--VIII in \figref{fig:figure1}a). Then we show how a linear bend transforms into a kink when decorated by H along the most reactive (most curved) line (\figref{fig:figure1}b), and present the electronic transmission through a single kink in \secref{sec:singlekink}. The kink acts as an effective barrier with transmission depending on kink-angle, $\varphi$. In \secref{sec:twokinks} we study how two parallel kinks lead to the formation of a pseudo-ribbon type electronic structure. Finally, in \secref{sec:multiplekinks} we demonstrate the opening of a transport gap for multi-kink systems, such as the one shown in \figref{fig:figure1}b, and \secref{sec:conclusions} gives our conclusions.

\section{\label{sec:setup}System setup}

The bend we consider in \figref{fig:figure1}a is created along the armchair direction by fixing the first and last six rows of carbon atoms and shortening the separation $L$, while the rest of the atoms are allowed to relax. A separation of {$L=90$~\AA} is chosen in order to obtain realistic curvatures\cite{LoPeTe.2012}. We have first assessed the reactivity of the structure at positions with different local curvature, see positions I--VIII in \figref{fig:figure1}a. Subsequently we have relaxed the structure where lines of hydrogen (\figref{fig:figure1}b) have been placed at the points of lowest radius of curvature, \ie the points of highest local reactivity. This particular system is meant to illustrate the potential of the hydrogen adsorption mechanism, and to gain insight into the modification of the electronic properties due to the hydrogen lines. In a corresponding experimental setup we can imagine placing graphene across a trench, which allows hydrogen adsorption on either side of the sheet.

The atomic and electronic structure calculations are based on density functional theory (DFT) using the SIESTA\cite{soler_siesta_2002} code, and the PBE-GGA exchange-correlation\cite{perdew_generalized_1996} functional. We employ periodic boundary conditions (PBC) in the direction along the bend with a cell-width of four carbon dimers, and 10 Monkhorst-Pack $k$-points. We use a mesh cut-off of 500~Ry throughout. When calculating reactivity in the form of reduced reaction barriers the unit-cell is chosen so that the distance between single hydrogen atoms is larger than 8.5~\AA. This ensures low hydrogen-hydrogen interaction, which is known to impact reaction barriers\cite{kerwin_sticking_2008}. In the total energy calculations of relaxed atomic geometries and reaction barriers we also use PBC transverse to the bend (5~$k$-points). We use a TZP basis-set for hydrogen and a SZ basis for carbon, except in the reaction barrier calculations where we compare calculations using a DZP and SZP basis for the four carbon atoms nearest to the hydrogen. In the barrier determinations we furthermore use spin-polarized calculations, because of unpaired electrons. For the relaxed geometries a force tolerance of {0.01~eV/\AA} is used, and the final energies are corrected for basis set superposition error (BSSE).

Based on the computed atomic and electronic structures we subsequently use the TranSIESTA\cite{BrMoOr.2002} method to calculate the electronic conductance per transverse-bend unit-cell width. To this end we attach semi-infinite flat graphene electrodes to each side of the selected kinks, e.g. replacing sections $S_1$ and $S_2$ in \figref{fig:figure1}b by semi-infinite electrodes in order to calculate the transmission through the single kink separating $S_1$ and $S_2$. In the conductance calculations we employ a dense transverse $k$-point grid of 400~points.

\section{\label{sec:results}Results}
\subsection{\label{sec:adsorption}Adsorption barrier}

\begin{figure}[tbp]
	\centering
	\includegraphics[width=.45\textwidth]{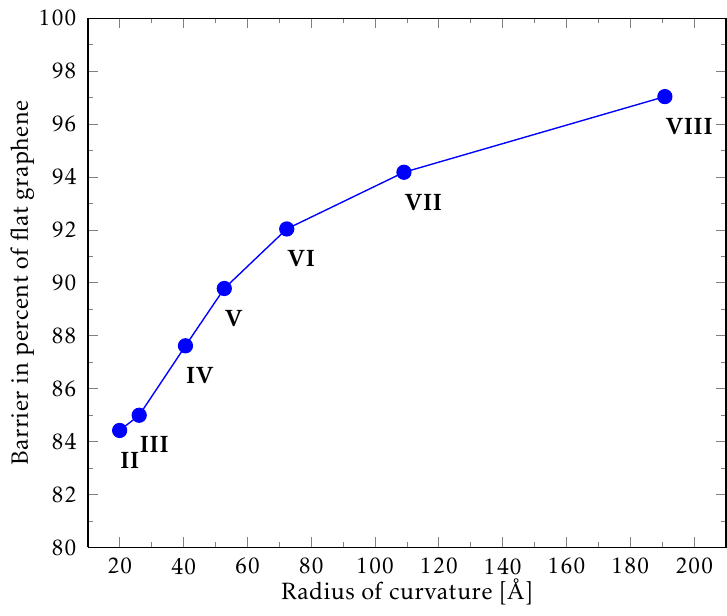}
	\caption{(Color online) Calculated reaction barriers for hydrogenation of bend graphene as a function of the local radius of curvature (II--VIII in \figref{fig:figure1}a). Flat graphene (position I) has an infinite radius of curvature and is used to normalize the barriers. Calculations are spin-polarised and allow for atomic relaxation.}
	\label{fig:figure2}
\end{figure}

Adsorption of hydrogen on graphene involves a reaction barrier which needs to be overcome before the single hydrogen atom sticks to the graphene sheet. Several investigations based on DFT calculations show that atomic hydrogen adsorbs on-top on flat graphene with a barrier about 0.2~eV and binding energy in the range of 0.7-1~eV\cite{jeloaica_dft_1999,sha_first-principles_2002,XiLiKr.11,hornekaer_clustering_2006}. Thus a minimum kinetic energy for the first hydrogen to react\cite{jeloaica_dft_1999,sha_first-principles_2002} is required in an out-of-equilibrium situation such as in an atomic beam\cite{XiLiKr.11}. Casolo \etal\cite{casolo_understanding_2009} calculated the reaction barrier and adsorption energy for multiple hydrogen atoms on flat graphene. In agreement with other studies they found decreased barriers to sticking for the second H atom, compared to the barrier for adsorbing a single H atom on a clean surface\cite{SlRaHo.09}.

Here, we first focus on the trends in the change in adsorption barrier as a function of the local curvature of the graphene sheet. To this end we have considered atomic hydrogen absorption at the on-top carbon positions at points with different curvature on the bent structure, see \figref{fig:figure1}a (positions I--VIII). The barrier is determined by calculating the total energy for each position of hydrogen above graphene as the hydrogen is moved successively closer to the graphene. Following the adsorption investigations on flat graphene by Ivanovskaya \etal\cite{ivanovskaya_hydrogen_2010} we perform a relaxation in each step of the hydrogen-bonded carbon atom and its three nearest neighbours. Using the method described above we obtain a reaction barrier of 0.22~eV on locally flat graphene. This is comparable to results obtained by several groups using DZP or plane wave basis sets and the PW91 functional\cite{jeloaica_dft_1999,sha_first-principles_2002,hornekaer_clustering_2006,kerwin_sticking_2008}. We find that a SZP basis set for the relaxed carbon atoms yields a reduced barrier height of 0.18~eV (both basis sets with orbital range corresponding to an energy-shift of 0.01~eV). Hence, we use the SZP basis in the following reaction barrier calculations in order to lower computation time.

For the positions (I-VIII) we obtain the reaction barrier for adsorption of hydrogen as a function of the local radius of curvature (RoC) shown in \figref{fig:figure2}. The second least curved position (VIII), resulting in a large RoC, reduces the barrier by roughly 3\% compared to flat graphene (position I). The most curved position (II) in our considered structure has a minimum RoC of {$\sim\!20$~\AA} resulting in a barrier reduction of roughly 16\%. For comparison, this RoC corresponds roughly to the radius of a (25,25) nanotube. Experiments show that \buck\ molecules are more reactive than CNTs, which in turn are more reactive than graphite\cite{rugrbi.2002}. Thus a significant increase in reactivity is expected for the graphene with the linear bend, and a simple Arrhenius estimate using our data yields a factor of 3--4 at room temperature (300~K). We have also performed calculations using the less rigorous DFTB method\cite{aradi_dftb_2007} and obtained results in agreement with the trend in reaction barrier reduction obtained above.

We may understand the reaction barrier and its change with curvature by considering the changes in carbon bond lengths. The barrier is due to the fact that the reacting carbon atom has to be pulled out of the graphene plane, stretching the strong carbon-carbon bonds, when reacting with the incoming hydrogen atom. When the graphene sheet is curved the carbon atom is already slightly out of the plane, and thus the energy required to pull the atom further out of plane is decreased compared to flat graphene. The ortho- and para-locations in the graphene hexagon have been shown to be the preferred locations for  hydrogen adsorption in studies of flat graphene\cite{hornekaer_clustering_2006,kerwin_sticking_2008,sljivancanin_structure_2011}. With this in mind as well as the curvature-related reduction of reaction barriers we conclude that the considered system allows the adsorption of hydrogen atoms in single lines along armchair-edges. The kink in the atomic structure due to the $sp^3$-binding of a single H makes the graphene curve even more in its vicinity which, in turn, preferentially lowers the barrier for absorption of a H along the linear bend. This suggests a mechanism where the hydrogen adsorption is propagating and leads to the decoration of the entire linear bend turning it into a kink-line. It may be viewed as analogous to crack formation mechanisms, where the breaking of a bond increases the stress on neighbouring bonds - only in this case, the graphene is hydrogenated rather than broken or destroyed.

\subsection{\label{sec:singlekink}Single kink}

\begin{figure}[tbp]
	\centering
	\includegraphics[width=.45\textwidth]{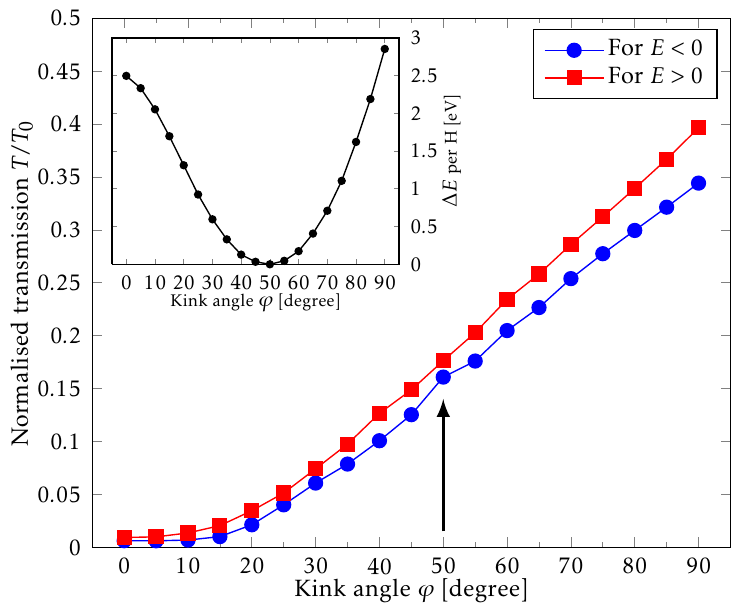}
	\caption{(Color online) Electronic transmission through a single kink normalized with the pristine graphene
transmission ($T_0$) as a function of the kink angle, $\varphi$, for electrons ($E\!>\!0$) and holes ($E\!<\!0$).
The arrow indicates the normalised transmission at the equilibrium angle determined from the total energy calculations shown in the inset.}
	\label{fig:figure3}
\end{figure}

Next, we examine the energetic and transport properties of kink-lines in the armchair-direction. We first consider a single kink with angle $\varphi$, \eg between sections $S_1$ and $S_2$ in \figref{fig:figure1}b. The kink-angle $\varphi$ is varied in the range \mbox{$0^\circ\!\ldots 90^\circ$}, while the three nearest unit cells on each side of the kink are allowed to relax. The total energy per H is shown as a function of $\varphi$ in the inset of \figref{fig:figure3}, showing a minimum energy for \mbox{$\varphi\!\approx\!50^\circ$}. This angle roughly corresponds to the angle in a $sp^3$-configuration where \mbox{$2\varphi\!=\!109.5^\circ$}. The adsorption of H causes local changes in the geometry, \ie only the carbon atoms very close to the kink are moved, while the remaining structure remains unperturbed. For this reason, the adsorption of hydrogen atoms can be considered as a process which locally pins the bend.

The electron transmission per unit-cell width is linear in energy for pristine graphene in a energy range around the charge neutrality point ($E\!=\!0$), \eg \mbox{$T_0\!\propto\!E$}. We find similarly that the calculated kink-transmission curves also are linear, and therefore we express the results for the transmissions in terms of the roughly energy independent ratio \mbox{$T/T_0\!=\!\text{const}$}. The kink breaks the electron-hole symmetry and we fit \mbox{$E\!>\!0$} and \mbox{$E\!<\!0$} separately, as shown in \figref{fig:figure3}. Larger kink-angles result in an increase in the overall transmission, which may be contributed to a better $\pi$-orbital overlap across the kink. For the equilibrium angle, \mbox{$\varphi\!=\!50^\circ$}, the ratio $T/T_0$ is close to $0.17$ in both regions (indicated by the arrow in \figref{fig:figure3}), corresponding to a transmission reduction of 83\%. Thus, we see that the hydrogen-induced kinks in graphene can be used to form effective electron barriers. We now turn to the effect of multiple barriers and periodic kink structures in order to examine resonant tunnelling phenomena and band gap formation.

\subsection{\label{sec:twokinks}Two kinks}

\begin{figure*}[tbp]
	\centering
	\includegraphics[width=.9\textwidth]{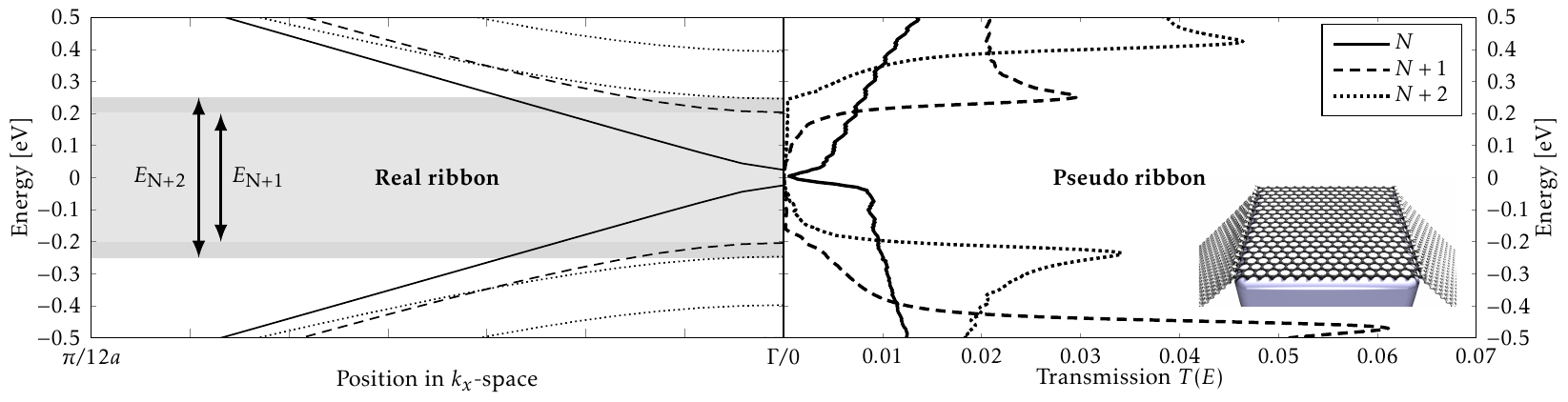}
	\caption{(Color online) (Left) Band structures for H-passivated armchair ribbons with varying width, $N$.
 The ribbons are a zero-gap semi-conductor, a semi-conductor with band gap \mbox{$E_\text{N+1}\!=\!0.4$~eV}, and a semi-conductor with band gap \mbox{$E_\text{N+2}\!=\!0.5$~eV} for widths $N,\,N+1,\,N+2$, respectively. All widths are based on the number of carbon atom lines \mbox{$N\!=\!17$}. The band gaps are indicated by arrows and highlighted. (Right) The electronic transmission functions for the corresponding pseudo-ribbons, \ie across two parallel kinks of varied separation as shown in inset.}
	\label{fig:figure4}
\end{figure*}

Band structure calculations show that periodic nano-rippling of the graphene is not sufficient to create a band gap\cite{WaZhLi.2011} due to the low scattering by the elastic deformation\cite{LoPeTe.2012}. In contrast, periodic arrangements of adsorbed hydrogen can indeed induce a semiconducting band gap\cite{chernozatonskii_two-dimensional_2007,WaZhLi.2011}. The electronic band structures of hydrogen lines on flat graphene have been examined by Chernozatonskii \etal\cite{chernozatonskii_two-dimensional_2007,chernozatonskii_nanoengineering_2010}, and recently also for nano-rippled graphene\cite{WaZhLi.2011}. Here we show how two parallel kinks lead to a local electronic structure which resembles that of an isolated GNR between the kinks. Such structures could be produced experimentally using the techniques described by Pan \etal\cite{pan_wrinkle_2011}. Hydrogen terminated armchair GNRs are semiconducting but have a small energy gap when the width in atomic lines is $N = 3L-1$, where $L$ is an integer\cite{BaHoSc.6}. In \figref{fig:figure4} we compare the electronic bandstructure for armchair GNR (aGNR) (left panel) to the electronic transmission through two kinks separated by the corresponding aGNR width (right panel). In the present case the initial width (or, kink separation) is {$N\!=\!17$} atomic lines of carbon, which shows a semi-metallic behaviour in the transmission spectrum with a small transport gap. In accordance with isolated aGNRs the next two widths {$N\!=\!18,\,19$} are semi-conducting, while the last investigated width {$N\!=\!20$} is semi-metallic again (not shown). The close correspondence between the electronic band structure for the GNR and the transmission gap for the double kink system allows us to consider the structure between two kinks as a pseudo-ribbon.

For the semi-conducting pseudo-ribbons transport gaps surrounded by van Hove-type 1D behaviour are seen in the transmission functions (\figref{fig:figure4}, right panel). The transport gaps, {$E_\text{gap}\!=\!0.4$~eV} and {$E_\text{gap}\!=\!0.5$~eV}, are in reasonable agreement with power-law scaling of $E_\text{gap}$ with width found for aGNRs\cite{BaHoSc.6}. We note that the pseudo-ribbon breaks electron-hole symmetry. For the $N=18$ case a larger van Hove resonance is seen at the valence band-edge, while for $N=19$ a larger resonance is seen at the conduction band-edge. There are small transmission values within the electronic band gap due to leakage through the barriers, which we expect to introduce shifts in the energies between the real and pseudo aGNR.

\subsection{\label{sec:multiplekinks}Multiple kinks}

\begin{figure*}[tbp]
	\centering
	\includegraphics[scale=1]{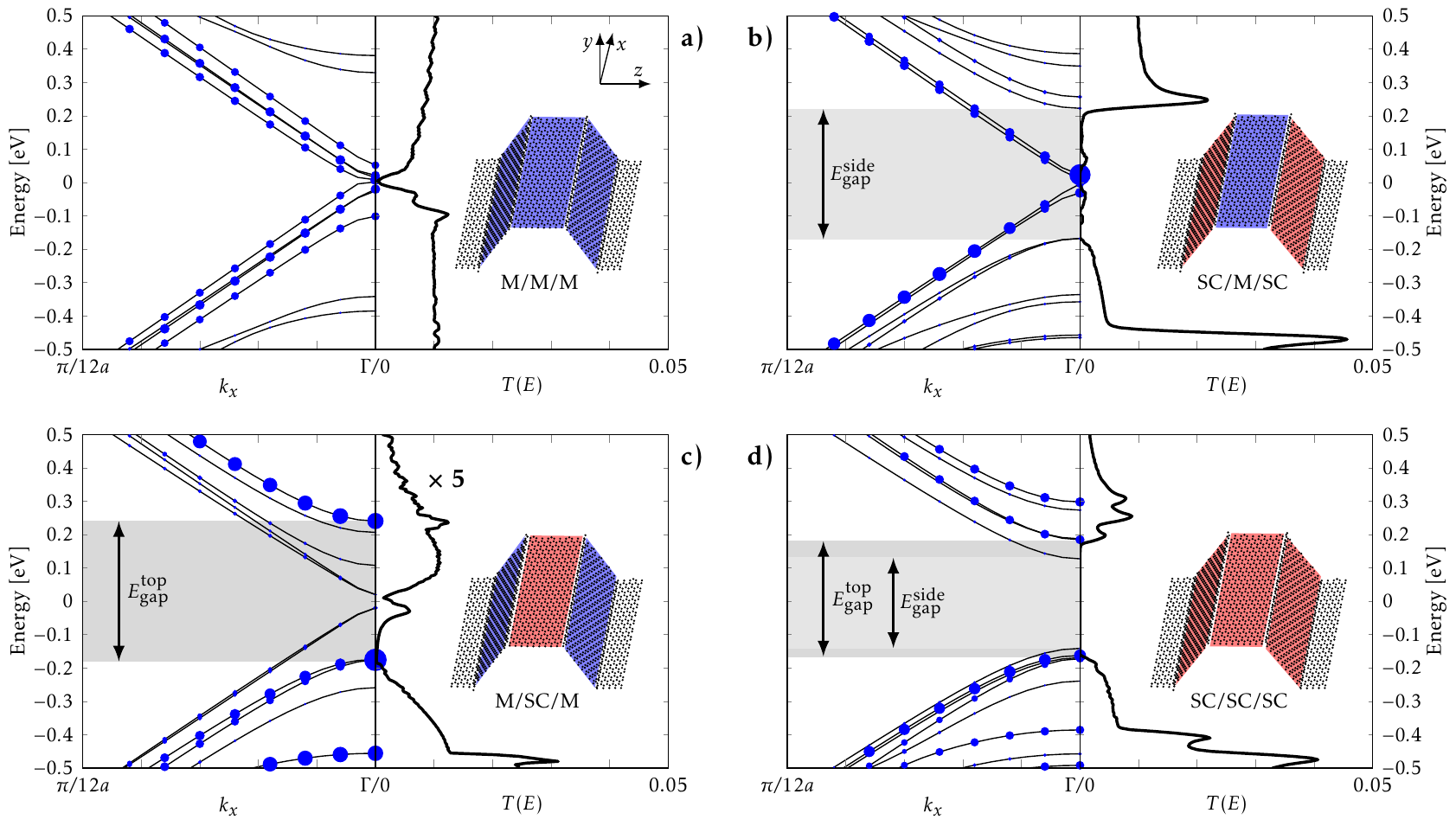}
	\caption{(Color online) Projected band structure and transmission through structures with multiple kinks. The top (section $S_3$) and connecting pseudo-ribbons (sections $S_2/S_4$) are varied in width, changing their electronic properties. The projected band structure along the pseudo-ribbons (\ie~$k_x$) is shown using filled circles with radii proportional to the weight on section $S_3$. Semi-metallic~(M) pseudo-ribbons corresponding to $N=17$ are shown in blue, while semi-conducting~(SC)  ribbons of width $N+1$ are shown in red. Bandgaps of the top sections $E^\text{top}_\text{gap}$ and connecting sections $E^\text{side}_\text{gap}$ are highlighted. These are $E^\text{side}_\text{gap}=0.39$~eV~(b), $E^\text{top}_\text{gap}=0.42$~eV~(c), and $E^\text{top}_\text{gap}=0.35$~eV, $E^\text{side}_\text{gap}=0.27$~eV~(d). The transmission $T_e(E)$ (per simple transverse unitcell) is determined across all kinks in the $z$-direction, and transmission gaps comparable to the bandgaps are observed in subfigures b and d.}
	\label{fig:figure5}
\end{figure*}

In order to illustrate the behaviour of systems with more kinks we consider a system consisting of four hydrogen-induced kinks, as illustrated in Figs.~\ref{fig:figure1}b and \ref{fig:figure5}. The sections $S_1$/$S_5$ are now replaced by left/right infinite lead structures, and the "top" $S_3$ pseudo-ribbon is connected to the leads via the "side" $S_2,S_4$ pseudo-ribbons. We keep $S_2,S_3$ identical for simplicity, and determine the transmission across the kinks (in the $z$-direction in \figref{fig:figure5}), which is experimentally more feasible. We now investigate how the different sections influence the total transport for the four possible combinations of semi-conducting (SC) and semi-metallic (M), corresponding to the pseudo-ribbon widths $N, N+1$ used in \figref{fig:figure4}.

In order to analyse the transmission we single out the band structure projected on to the top section, $S_3$ (excluding carbon and hydrogen atoms at the kink), in the band structure along the pseudo-ribbon direction. The weight on $S_3$ is represented by a circle of radius $R_{nk}$,
\begin{align}
	R_{nk} \propto \sum_{i \in S_3} \vert \Psi\left( n,k_x,i \right)\vert^2.
\end{align}
Here $\Psi$ is the wave function at $k_x$ ($k_z=0$) with $n$ ($i$) being the band (orbital) index. The obtained projected band structures are shown in the left parts of each subfigure in \figref{fig:figure5}. Generally, some bands have no weight (no circles), while others have significant weight indicating that there is little mixing between the orbitals from section $S_3$ and other sections.

In Figs.~\ref{fig:figure5}a and \ref{fig:figure5}b we consider pseudo-ribbons with semi-metallic top regions, namely $S_2/S_3/S_4$ being M/M/M, and SC/M/SC, respectively. For the all-metal pseudo-ribbons, M/M/M, an almost energy-independent transmission function is seen with a transmission close to that of the metallic double-kink in \secref{sec:twokinks}. The SC/M/SC structure shows a transport-gap similar to that of the single SC pseudo-ribbon with van Hove resonances, while the $S_3$-projected band structure reveals isolated metallic states within the gap. We note that the transmission at the resonances for the SC/M/SC structure is larger than the corresponding M/M/M transmission. For the case of semi-conducting top pseudo-ribbons in \figref{fig:figure5}c and \figref{fig:figure5}d, we note that M/SC/M show a greatly reduced transport gap compared to the single pseudo-ribbon case (also, note the scaling of the transmission axis), while the SC/SC/SC structure shows a complete extinction of the transmission in the electronic gap, as expected. Generally, we find that the main behaviour of the transmission is controlled by the connecting sections $S_2,S_4$, \ie there is a good correspondence between the side section bandgaps $E^\text{side}_\text{gap}$ and the transmission gaps.

\section{\label{sec:conclusions}Conclusions}

The presented investigations show that linear kink-line structures may form in graphene by reacting with atomic hydrogen along a linear bend in the sheet. The adsorption barrier is lowered in the close vicinity of the bend which can be exploited to form the kink. In particular, we have shown that a radius of curvature of {$\sim\!20$~\AA} reduces the hydrogen adsorption barrier by roughly 16\% compared to H adsorption on pristine graphene. The calculations suggest that once a single hydrogen atom has been adsorbed, the induced local kink and resulting increase in local curvature makes it easier for the following H to adsorb, thus creating a propagating kink formation. A full line of hydrogen atoms pins the structure and divides the electronic systems into different regions. We have shown that the electronic transmission through a single kink is reduced by 83\% compared to pristine graphene, meaning that the kink-line acts as an efficient barrier for electron motion. We have demonstrated how two close-by parallel kinks form a pseudo graphene nanoribbon with similar behaviour of the electronic structure as for isolated nano-ribbons. The transmission function displays transport gap features corresponding to the isolated nano-ribbon band gaps.

The present work thus suggests that it may be feasible to template functional electronic nanostructures using the conformation of graphene, \eg to the substrate, and that this in turn induce changes in local reactivity. Our work clearly calls for extensions in a number of directions. First of all more calculations are needed in order to investigate the kink-line propagation reaction proposed by our results. To this end it is important to include a realistic description of the actual substrate. It is also interesting to consider other adsorbate species, possibly introducing doping of the pseudo-ribbons and electronic gating. Finally, decoration and pinning of the edges of other geometries such as "bubbles" or "blisters" is of interest, \eg in order to produce GAL-like structures\cite{Pedersen2008} without perforating the graphene sheet.

\begin{acknowledgments}
We appreciate helpful discussion with Dr.~H.~Sevin\c{c}li.
We thank the Danish Center for Scientific Computing (DCSC) for providing computer resources.
The Center for Nanostructured Graphene is sponsored by the Danish National Research Foundation.
\end{acknowledgments}


\end{document}